\documentclass[aps,prb,onecolumn,showpacs,reprint]{revtex4-1}

\usepackage{wrapfig}
\usepackage[]{graphicx,xcolor}
\usepackage{tabularx}
\usepackage{booktabs}
\usepackage{textcomp}
\usepackage{amsmath}
\usepackage{amssymb}
\usepackage[]{graphics}
\usepackage{amssymb}
\usepackage{amsmath}
\usepackage{color}
\usepackage{ifpdf}
\newcommand{\executeiffilenewer}[3]{%
\ifnum\pdfstrcmp{\pdffilemoddate{#1}}%
{\pdffilemoddate{#2}} > 0 {\immediate\write18{#3}}\fi}

\ifpdf\usepackage{epstopdf}\fi
\begin{document}

\title{Plasmon-polariton induced modification of silicon nanocrystals photoluminescence in presence of gold nanostripes}

\author{S. A. Dyakov}
\email{e-mail: s.dyakov@skoltech.ru}
\affiliation{Skolkovo Institute of Science and Technology, 143025 Moscow, Russia}

\author{D. M.~Zhigunov}
\affiliation{Faculty of Physics, Lomonosov Moscow State University, 119991 Moscow, Russia}

\author{A.~Marinins}
\affiliation{School of Engineering, KTH Royal Institute of Technology, Stockholm, Sweden}

\author{O.~A.~Shalygina}
\affiliation{Faculty of Physics, Lomonosov Moscow State University, 119991 Moscow, Russia}

\author{P.~P.~Vabishchevich}
\affiliation{Faculty of Physics, Lomonosov Moscow State University, 119991 Moscow, Russia}

\author{M.~R.~Shcherbakov}
\affiliation{Faculty of Physics, Lomonosov Moscow State University, 119991 Moscow, Russia}

\author{D.~E.~Presnov}
\affiliation{Skobeltsyn Institute of Nuclear Physics, 119991 Moscow, Russia}
\affiliation{Faculty of Physics, Lomonosov Moscow State University, 119991 Moscow, Russia}

\author{A.~A.~Fedyanin}
\affiliation{Faculty of Physics, Lomonosov Moscow State University, 119991 Moscow, Russia}

\author{P.~K.~Kashkarov}
\affiliation{Faculty of Physics, Lomonosov Moscow State University, 119991 Moscow, Russia}
\affiliation{National Research Centre Institute, 123182 Moscow, Russia}

\author{S.~Popov}
\affiliation{School of Engineering, KTH Royal Institute of Technology, Stockholm, Sweden}

\author{S.~G.~Tikhodeev}
\affiliation{A.~M.~Prokhorov General Physics Institute, 119991 Moscow, Russia}
\affiliation{Faculty of Physics, Lomonosov Moscow State University, 119991 Moscow, Russia}

\author{N.~A.~Gippius}
\affiliation{Skolkovo Institute of Science and Technology, 143025 Moscow, Russia}

\date{\today}

\begin{abstract}
We report the results of theoretical and experimental studies of photoluminescence of silicon nanocrystals in the proximity of plasmonic modes of different types. In our samples, the type of plasmonic mode is determined by the filling ratio of a one-dimensional gold grating which covers the thin film with silicon nanocrystals on a quartz substrate. We analyze the extinction and photoluminesce spectra of silicon nanocrystals and show that the emitted light is coupled to the corresponding plasmonic mode. We also demonstrate the modification of the extinction and photoluminesce spectra under the transition from surface plasmon-polaritons to waveguide plasmon-polaritons with the decrease of the gold filling ratio from 1 to 0.35. Finally, we analyze the contribution of individual silicon nanocrystals to the overall photoluminescence intensity. We conclude that silicon nanocrystals ensemble can be broken down into optically bright and optically dark nanocrystals.  The experimental extinction and photoluminescence spectra are in good agreement with theoretical calculations performed by the Fourier modal method in the scattering matrix form.
\end{abstract}


\maketitle 
It is well known that a quantum dot placed in a strongly non-homogeneous dielectric environment can exhibit optical properties that are quite different from those of a quantum dot in free space \cite{noda2007spontaneous, holland1985waveguide, guzatov2012plasmonic}. From the optical point of view, the emission intensity depends on the excitation efficiency, out-coupling efficiency as well as on the probability of a quantum dot to radiate photons or to transfer its energy to the matrix nonradiatively. These parameters can be effectively tuned in the presence of metals. The effect of the influence of metallic nanostructures on the emission characteristics of nearby emitters have been observed in a variety of different molecules and quantum dots \cite{guzatov2012plasmonic, tam2007plasmonic, muskens2007strong, PhysRevLett.101.116801}. In particular, it has been shown that the photoluminescence intensity of silicon quantum dots can be effectively enhanced in the proximity of metallic structures which support surface plasmon-polariton modes \cite{Takeda2006c} or localized surface plasmon modes \cite{PhysRevB.93.205413, goffard2013plasmonic, Biteen2007, sugimoto2015plasmon, gardelis2016twenty, nychyporuk2011strong, inoue2015surface, harun2013gold}. This fact is important from the viewpoint of potential optoelectronic applications of silicon nanocrystals given that silicon nanocrystals are CMOS-compatible and exhibit room-temperature photoluminescence \cite{zhigunov2009effect}. 

Despite the number of publications devoted to the optical properties of silicon nanocrystals in the proximity of metals, it is interesting to study the photoluminescence in a plasmonic system where both types of plasmonic modes could exist. In this paper, we focus on theoretical and experimental studies of the optical properties of silicon quantum dots in the proximity of one-dimensional arrays of gold nanostripes. Depending on the filling ratio, the gold nanostripes array supports either localized surface plasmons or propagating surface plasmon-polaritons. As shown in Ref.\,\citenum{PhysRevB.93.205413} for narrow gold nanostripes, in the presence of the waveguide containing silicon nanocrystals, the localized surface plasmons are strongly coupled to the quasiguided modes resulting in the formation of a waveguide plasmon-polariton. Here we aim to demonstrate the modification of emission characteristics of the silicon nanocrystals under the transition from surface plasmon-polariton \cite{porto1999transmission} to waveguide plasmon-polariton \cite{Christ2003b, Christ2004, PhysRevB.93.205413}.

\begin{figure*}[t!]
\centering
\includegraphics[width=1\textwidth]{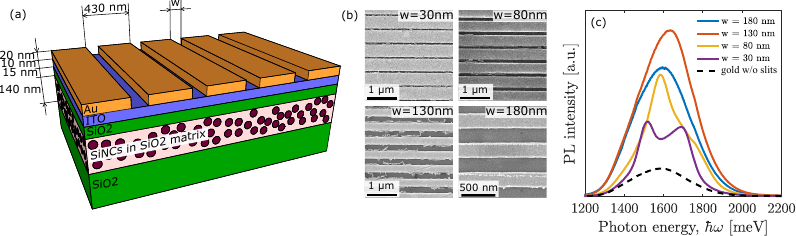}
\caption{(Color online) (a) Schematic view of the sample with silicon nanocrystals and a gold grating. Silicon nanocrystals are shown by circles. (b) SEM images of  structures under study. (c) Experimental PL spectra for different air slits widths. The black dashed line shows the spectrum for the uniform structure without slits.}
\label{sam}
\end{figure*}

The structure of the paper is as follows. First, we describe the fabrication technique of our structures with silicon nanocrystals and the details of the experimental setup. We give a brief description of the theoretical method for the PL intensity calculation. Then, we present the results of extinction and photoluminescence measurements as well as the results of numerical simulations. We show that the extinction and photoluminescence spectra have features associated with either surface plasmon-polaritons or waveguide plasmon-polaritons depending on the air slit width. We also calculate the electromagnetic near-field distribution of the light which is coming in to the sample from free space. Finally, we analyze the contribution of individual silicon nanocrystals to the overall PL intensity.

%%%%%%%%%%%%%%%%%%%%%%%%%%%%%%%%%%%%%%%%%%%%%%%%%%%%%%%%%%
%\section{Methods}
%\label{sec:methods}
%\subsection{Structure and experimental details}
%\label{subsec:exp}

The schematic of the investigated structure is shown in Fig.\,\ref{sam}. The structure consists of a periodic array of air slits in a 20\,nm thick gold film deposited on a quartz substrate covered by thin SiO$_2$ film with silicon nanocrystals. The air slits width was varied between $w=$\,30\,nm and 180\,nm with a step of 50\,nm. The pitch size was kept fixed at 430\,nm. Silicon nanocrystals were evenly distributed in the SiO$_2$ film on the depths from 15\,nm to 155\,nm. Scanning electron microscopy (SEM) images of the sample with different air slit widths are shown in Fig.\,\ref{sam}b. The images were captured by a high-resolution Field Emission Scanning Electron Microscope Supra 40 (Carl Zeiss).

Reactive evaporation of SiO powder in an oxygen atmosphere was used to deposit SiO$_x$ ($x\approx 1.7$) films on quartz substrates. Films thickness was equal to about 140 nm, while a capping 15\,nm thick SiO$_2$ layers were also deposited on a top of each film by increasing an oxygen pressure during evaporation. After the deposition the conventional tube furnace annealing at 1100$^\circ$C for 1 hour in N$_2$  atmosphere was used in order to fabricate Si nanocrystals in SiO$_2$ matrix (see for details Ref.\,\citenum{zhigunov2012photoluminescence}).

For the gold nanostripe fabrication, the glass substrate with silicon nanocrystals is covered with 10-nm thick indium tin oxide (ITO) layer as an adhesion promoter between gold and silica. Then, the sample is cleaned, CSAR 62 resist spin coated and baked forming a 140\,nm thick uniform layer. Next, 500$\times$500$\,\mu$m grating is patterned with electron beam lithography system (Raith 150, 25\,kV acceleration voltage) using fixed beam moving stage (FBMS) mode. This allows uniform exposure. After development 20\,nm of gold are deposited in high vacuum e-gun evaporation system (Eurovac). Then, the lift-off process is performed by immersing the sample in acetone. This removes the photoresist with the excess Au leaving only Au grating lines deposited on ITO, which serves as a transparent adhesion promoter.

Transmittance spectra were measured as a function of the angle of light incidence. In the setup, light from a broadband source (50\,W halogen lamp) is collimated and slightly focused to a spot of ca 500 $\mu$m in diameter. The polarization state is controlled by a Glan-Taylor polarizer. The transmitted beam is collected and sent to a compact CCD-based visible spectrometer. The sample is held by a 3-axis holder that allows for the control of the incidence angle with a step of 1$^\circ$. The spectra were measured consecutively for the sample area and the substrate without gold grating; then, the sample spectra are normalized over the substrate spectra. 

Photoluminescence (PL) spectra were registered under the 325\,nm HeCd laser line excitation using 500 mm single-grating spectrometer equipped with an air-cooled CCD camera. The spectra were taken at room temperature and were corrected for the system response. 

%%%%%%%%%%%%%%%%%%%%%%%%%%%%%%%%%%%%%%%%%%%%%%%%%%%%%%%%%%%%%%%%%
%\subsection{Calculation of photoluminescence intensity}
%\label{subsec:calc}
%For the theoretical study of optical properties of the structures described above we calculate their extinction spectra and emissivity spectra.

The photoluminescence intensity was calculated as a power emitted by the oscillating electric dipoles uniformly distributed over the layer with silicon nanocrystals. From the population dynamics equations for silicon nanocrystals (see, for example, Ref.\,\citenum{biteen2005enhanced}) it follows, that in the approximation of low excitation power, the emission intensity of single dipole is proportional to the product of excitation efficiency $C_{exc}$ and the out-coupling efficiency $C_{pl}$ and inversely proportional to the total decay rate of the silicon nanocrystal $\Gamma$:
\begin{equation}
I_{i}\propto \frac{C_{exc} C_{pl}}{\Gamma}.
\label{eq_ccg}
\end{equation}
The parameters $C_{exc}$ and $C_{pl}$ are large in the vicinity of plasmonic modes due to the field enhancement. The recombination rate is determined by the number of resonances and increase in the near-field of metal due to the contribution of evanescent modes.

The overall PL intensity accounts for the contribution from all dipoles:
\begin{equation}
I = \sum_i I_i,
\label{eqsum}
\end{equation}
where emission intensity of $i$-th SiNCs in the ensemble is given by the formula (\ref{eq_ccg}) for the general case. In our samples, the closest distance between the silicon nanocrystals and metallic grating is 20\,nm which is rather large for plasmonic modes to notably influence the recombination rate of emitters \cite{guzatov2012plasmonic}. Hence we can assume the denominator in formula (\ref{eq_ccg}) to be roughly constant. Therefore, the contribution of each dipole, $I_i$, can be calculated as 
\begin{equation}
I_i \sim \left|\bold{E}\left(\hbar\omega_{exc},\bold{k}_{\parallel exc},\bold{r}_i\right)\right|^2\times\left|\bold{E}\left(\hbar\omega_{pl},\bold{k}_{\parallel pl},\bold{r}_i\right)\right|^2,
\label{eqprod}
\end{equation}
where $\bold{E}$ is the electric vector of incidence plane electromagnetic wave calculated at the photon energy $\hbar\omega_\alpha$, the in-plane projection of the photon quasimomentum vector $\bold{k}_{\parallel \alpha}\equiv\left(k_{x\alpha},k_{y\alpha}\right)$ and the coordinate of oscillating dipole $\bold{r}_i\equiv(x_i, z_i)$. The symbol $\alpha = $ "$exc$" or "$pl$" relates to the excitation or photoluminescence. The first factor in Eq.\,(\ref{eqprod}) is the excitation efficiency $C_{exc}$; it is proportional to the volume density of excited nanocrystals at the position $\bold{r}_i$. The second factor in Eq.\,(\ref{eqprod}), in accordance with the electrodynamic reciprocity principle has a meaning of an out-coupling efficiency $C_{pl}$ which is proportional to the probability for the emitted photon to come out from the sample and couple to the far field. 

Calculations of the electric field $\bold{E}$ are performed using the rigorous coupled wave analysis (RCWA) in the scattering matrix form \cite{Tikhodeev2002b, whittaker99, moharam1995formulation}. The general idea of this method is the Fourier decomposition of the electromagnetic field into planar waves with different projections of the momentum vector onto the direction of periodicity. In order to achieve a better convergence with respect to the number of plane waves, we employ the factorization rules \cite{li1996use}.

%%%%%%%%%%%%%%%%%%%%%%%%%%%%%%%%%%%%%%%%%%%%%%%%%%%%%%%%%%%%%%
%\section{Results and discussions}
%\label{sec:results}

The experimental PL spectra of the samples with different air slit widths are shown in Fig.\,\ref{sam}c. One can see that for all $w$ the PL intensity is higher than in the case of the sample without air slits. The PL spectrum at $w=30$\,nm has two peaks. With the increase of the air slit width, the higher energy peak disappears. To understand the above behaviour of the PL spectra with the increase of the air slit width, let us consider the angle-resolved extinction and PL spectra.

%Since the spectra shown in Fig.\,\ref{spectra}a were collected without a collimator, a wide range of in-plane wavenumber of PL light contributes to the overall PL signal. 

\begin{figure}[t!]
\centering
\includegraphics[width=0.8\columnwidth]{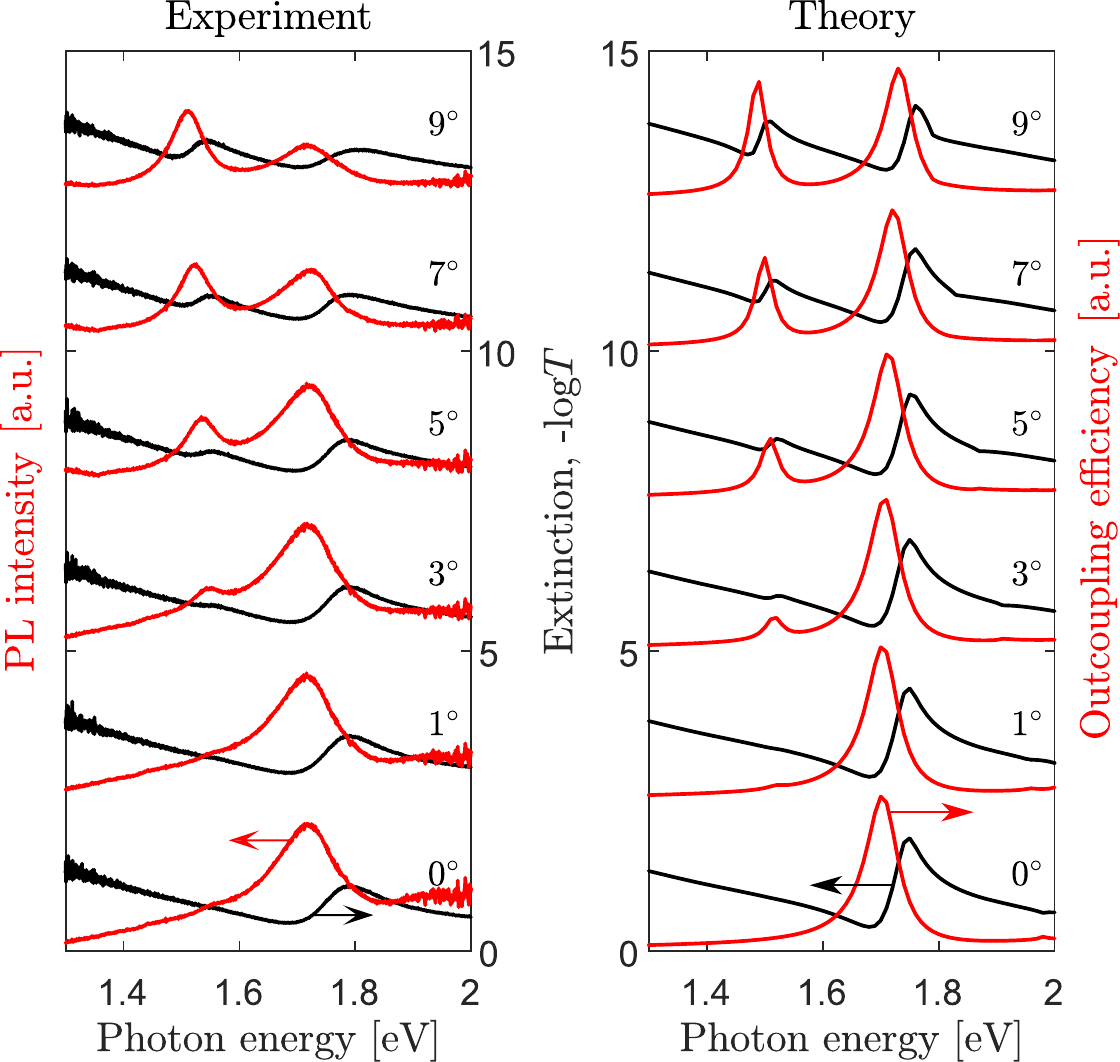}
\caption{(Color online) (a) Experimental extinction (black lines)
and PL spectra (red lines). The displayed angles denote the angle of light incidence. (b) Theoretical extinction (black lines) and out-coupling efficiency spectra (red lines). The displayed angles denote the angle of PL collection. Curves in panels (a) and (b) are shown for the sample with 30-nm-wide air slits.}
\label{sp}
\end{figure}

Angle-resolved TM-polarized experimental extinction and PL spectra, as well as their theoretical counterparts for the sample with the 30-nm-wide air slits are shown in Fig\,\ref{sp}a. Several important features in the peaks behaviour can be seen in Fig.\,\ref{sp}:
\begin{enumerate}
\item At $\theta=0$, both extinction and PL spectra have one peak. With the increase of $\theta$ this peak slowly shifts to higher energies.
\item With the increase of angle $\theta$ a new peak arise at the lower-energy side of the main peak. This peak shifts to lower energies.
\item Spectral position of PL peaks is located between extinction minima and extinction maxima.  
\end{enumerate}
The theoretical extinction and out-coupling efficiency spectra (Fig.\,\ref{sp}b) agree with the experimental results. The comparison between the experimental PL spectra and theoretical out-coupling efficiency reveals that the excitation efficiency does not play a significant role for the spectral position of the emission maxima.

\begin{figure}[t!]
\centering
\includegraphics[width=0.8\columnwidth]{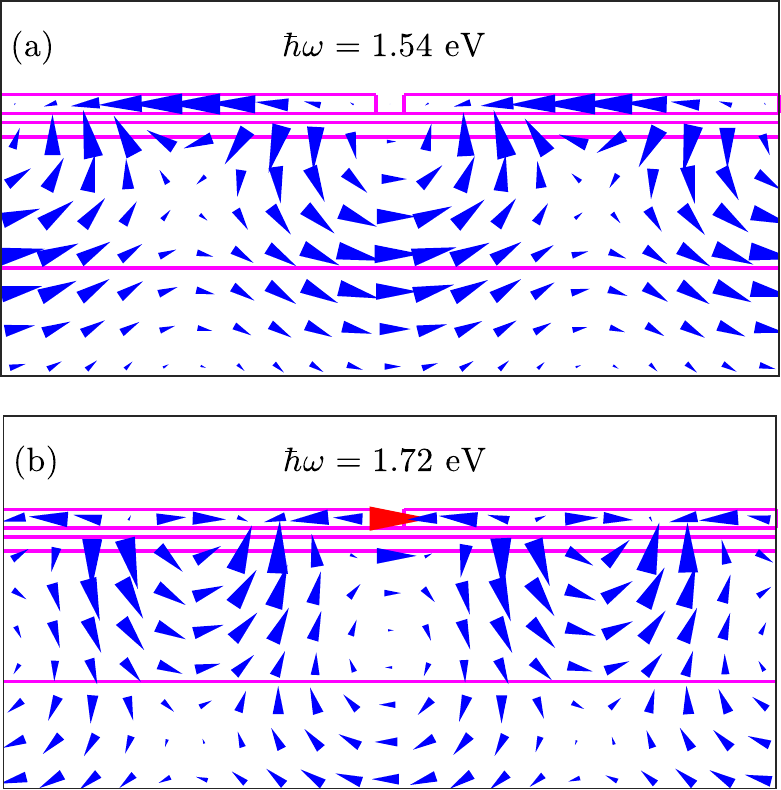}
\caption{(Color online) Calculated spatial distributions of the electric field in the structure with silicon nanocrystals for normal incidence of TM polarized light. The fields are shown for the photon energy of (a) $\hbar\omega=$\,1.54\,eV, (b) $\hbar\omega=$\,1.72\,eV. The size of triangles is proportional to the field at the central point of each triangle. Triangles specify the corresponding electric field direction by their orientation. The fields are depicted at the instant time when the field intensities, integrated over the displayed cross sections, reach a maximum.}
\label{field}
\end{figure}

The above resonances are attributed to surface plasmon-polariton modes \cite{porto1999transmission}. The difference of the upper and lower branches of the resonance from Fig.\,\ref{sp}a can be understood by inspecting the electric near-field distributions. The calculated spatial electric field distributions of the incident plane wave  is shown in Fig.\,\ref{field} for two photon energies, $\hbar\omega=1.54$\,eV and $\hbar\omega=1.72$\,eV which correspond to the theoretical out-coupling efficiency maxima at the polar angle $\theta=0.1^\circ$ and azimuthal angle $\phi=0$. In Fig.\,\ref{field}, the size of the triangles is proportional to the field strength at the center of each triangle. The length of the blue triangles is scaled to the amplitude of the incoming wave in a vacuum; in the case of red triangles, it is reduced by a factor of three, to prevent the triangles overlap. It can be seen from Fig.\,\ref{field} that for both photon energies the displayed field takes the shape of vortices and decays into the substrate. The electric fields at $\hbar\omega=1.54$\,eV and $\hbar\omega=1.72$\,eV represent the antisymmetric and symmetric propagating surface plasmon-polariton modes. At $k_x=0$ the lower antisymmetric mode is optically inactive and can only be observed in extinction spectra under an inclined incidence. 

\begin{figure*}[t!]
\centering
\includegraphics[width=1\textwidth]{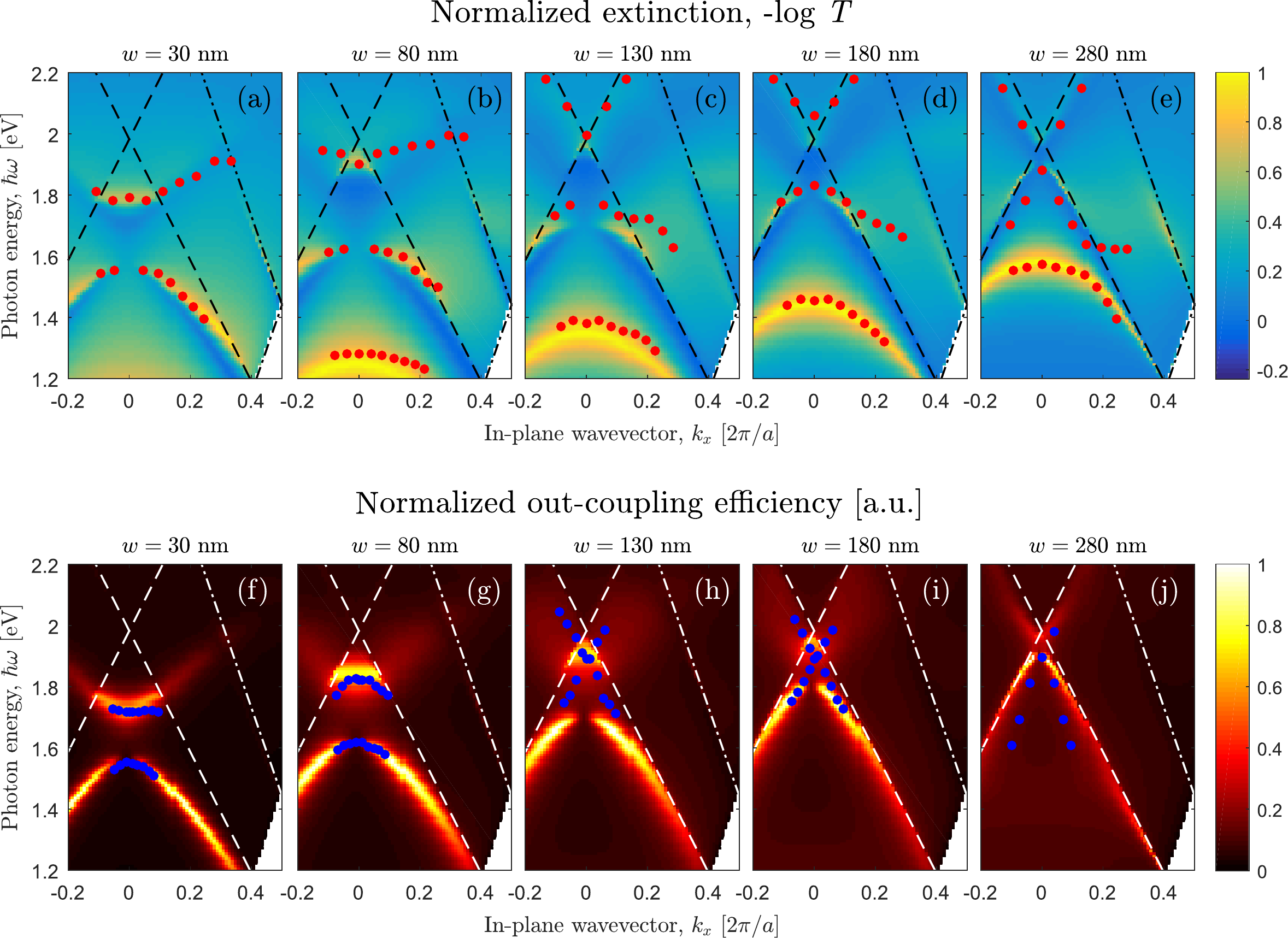}
\caption{(Color online) (a) Calculated $k_x$ and energy dependencies of the extinction (panels (a)--(e)) and emissivity (panels (f)--(j)) spectra of the TM-polarized light for the sample with different air slit width. The magnitude of the extinction coefficient and emissivity is shown by different colors and the color scale is explained on the right. The red dots in panels (a)--(e) represent the experimentally observed extinction peaks, while the blue dots in panels (f)--(j) show the experimental PL peaks. The peak positions of the sample with 280\,nm air slit are taken from Ref.\,\citenum{PhysRevB.93.205413}.}
\label{disp}
\end{figure*}

To understand the distinction between the experimental PL spectra for different air slit widths, let us consider the dispersions of extinction and photoluminescence for different air slit widths. For this purpose, we calculate the in-plane wavevector and energy dependence of the extinction and out-coupling efficiency for $w$ changing from 30\,nm to 280\,nm (Fig.\,\ref{disp}a--j). As it was already mentioned, the smallest air slit width of $w=30$\,nm corresponds to the propagating surface plasmon-polariton which is represented by two branches. The dispersion of these branches is not clearly seen in Fig.\,\ref{sp} angles are considered in PL measurements are rather small constrained by experimental setup. In the opposite case, when $w=280$\,nm, the hybrid mode of waveguide plasmon-polariton appear. A detailed analysis of this mode is carried out in Refs.\,\citenum{Christ2003b, Christ2004, PhysRevB.93.205413}. In the most general case, the series of graphs in (Fig.\,\ref{disp}a--j) demonstrates the transition from the propagating surface plasmon-polariton to the waveguide plasmon-polariton. Indeed, with the increase of the air slit width, the antisymmetric mode is transformed from the lower branch of the surface plasmon-polariton to the upper branch of the waveguide plasmon-polariton. The symmetric mode is transformed from the upper branch of surface plasmon to the lower branch of waveguide plasmon-polariton \cite{PhysRevB.93.205413}. The out-coupling efficiency spectra have two sets of peaks. The spectral positions of the out-coupling efficiency peaks are located close to those of the extinction minima. The experimental extinction and PL peaks positions are well described by our theoretical model as shown in Fig.\,\ref{disp} by circles. It should be noted that the lowest energy extinction mode is not seen in the photoluminesce spectra due to the high absorption in gold.

%%%%%%%%%%%%%%%%%%%%%%%%%%%%%%%%%%%%%%%%%%%%%%%%%%%%%%%%%%%%%%%%%%%%%%%%%%%

Up to now, we have been calculating the out-coupling efficiency of the samples under study as an integral over all emitter positions within the layer with silicon nanocrystals. At the same time, it is obvious that silicon nanocrystals, when to emit light, are in different optical conditions. From the viewpoint of expression for the overall PL intensity (\ref{eqsum}), it means that the excitation efficiency $\left|\bold{E}\left(\hbar\omega_{exc},\bold{k}_{\parallel exc},\bold{r}_i\right)\right|^2$ as well as the out-coupling efficiency $\left|\bold{E}\left(\hbar\omega_{pl},\bold{k}_{\parallel pl},\bold{r}_i\right)\right|^2$ depend on the emitter position $\bold{r}_i$. The spatial non-uniformity of the excitation efficiency indicates that the concentration of excited silicon nanocrystals in one part of the sample is higher than in the other. The spatial non-uniformity of out-coupling efficiency can be interpreted as following: the probability of an emitted photon to come out from the sample and to couple to the far field depends on the emitter position. As a result, silicon nanocrystals that are located in different positions within the active layer, give the different contribution to the overall PL intensity. 

\begin{figure*}[t!]
\centering
\includegraphics[width=1\textwidth]{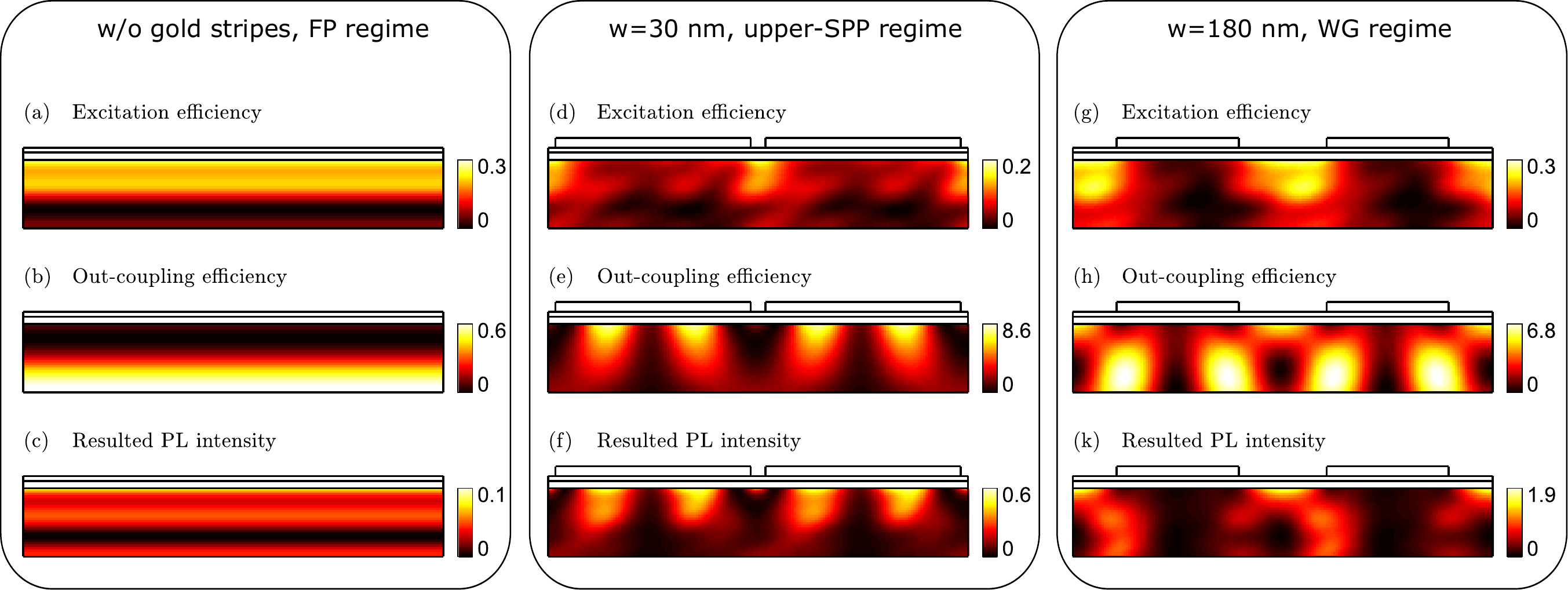}
\caption{(Color online) Calculated excitation efficiency (a), (d), outcoupling efficiency (b), (e) and resulted PL intensity as a function of dipole position within the layer with silicon nanocrystals. The color scales of the corresponding quantities are shown on the right. For comparison, the above quantities are shown for $w=30$\,nm (left panels) and $w=180$\,nm (right panels).}
\label{contrib}
\end{figure*}

The excitation efficiency and the outcoupling efficiency as a function of the emitter position is shown in Fig.\,\ref{contrib} for three different regimes. In the below discussion, all the structures are exposed by 325\,nm laser at $\theta_{exc}=45^\circ$ angle of incidence, a typical excitation scheme in our experimental setup. We start our discussion from the reference structure that has no gold layer. The photoluminescence is detected on silicon nanocrystals PL peak photon energy of 1.6\,eV at the normal collection angle. It can be seen from Fig.\,\ref{contrib}a that the excitation field is mainly localized in the sub-surface region causing the inhomogeneous profile of the excited silicon nanocrystals concentration. The spatial dependence of the outcoupling efficiency in the structure with slits is determined by the Fabry-Perot modes and as shown in Fig.\,\ref{contrib}b. Notably, for this particular structure and experimental conditions, the highest probability for the emitted photons to escape the structure is reached deep inside the emitting layer. The resulted PL intensity is found as a product of excitation efficiency and outcoupling efficiency and is displayed in Fig.\,\ref{contrib}c. It can be seen that in this structure, the excitation and outcoupling efficiency maps have a little overlap which leads to moderately low overall PL intensity. By changing the thicknesses of layers one can design the structure in such a way that excitation and out-coupling profiles match  each other yielding in higher PL intensity \cite{dyakov2012enhancement}. 

Let us consider the 30-nm-width air slit structure. The photoluminescence is detected at a  photon energy of 1.72\,eV at the normal collection angle, which corresponds to the upper surface plasmon-polariton mode (see Fig.\,\ref{disp}a). It can be seen from Fig.\,\ref{contrib}d that the excitation field is mainly settled close to the air slits. The outcoupling efficiency is determined by the electric field distribution of the upper surface plasmon-polariton mode and is shown in Fig.\,\ref{contrib}e. Since this mode has a standing-wave character, the outcoupling modulation coefficient, i.e. the ratio of the minimal and the maximal outcoupling efficiency within the emitting layer, is rather high. The resulted PL intensity for the 30-nm-width air slit structure is shown in Fig.\,\ref{contrib}f. The displayed profile of the PL intensity suggests that the major contribution to the overall PL intensity is brought by the nanocrystals located under the gold stripes in accordance with upper plasmon-polariton mode symmetry.

Finally, we move to the 180-nm-width air slit structure. The excitation field is maximal in the regions underneath the slits (Fig.\,\ref{contrib}g). As shown in Ref.\,\citenum{PhysRevB.93.205413}, the photoluminesce of such structure can be enhanced due to quasiguided modes. In the calculation, we choose the photoluminesce photon energy to match the upper quasiguided mode at the normal collection angle ($\hbar\omega=1.96$\,eV). The Fig.\,\ref{contrib}h reveals that the outcoupling efficiency takes the shape of loops which is dictated by the field distribution of quisiguided modes \cite{PhysRevB.93.205413}. The outcoupling modulation coefficient for this structure is very high ($\approx550$) which indicates the strongly non-uniform distribution of silicon nanocrystals contribution to the overall PL intensity. 

The above distributions over the ensemble of silicon nanocrystals account for purely optical effects and may be smoothed in reality due to the exciton migration process. Nevertheless, the very high values of the outcoupling modulation coefficients enable us to conventionally break down all silicon nanocrystals into optically bright and optically dark.

%%%%%%%%%%%%%%%%%%%%%%%%%%%%%%%%%%%%%%%%%%%%%%%%%%%%%%%%%%%%%%

%\section{Conclusion}
%\label{sub:conclusions}
In conclusion, we have theoretically and experimentally studied the optical properties of silicon nanocrystals covered by periodic arrays of plasmonic stripes. We have shown that the extinction and photoluminescence spectra have several sets of peaks, which are attributed to surface plasmon-polariton mode or waveguide plasmon-polariton mode depending on the air slit width. We also have demonstrated the transition between these two modes with the increase of the air slit width. Finally, we have analyzed how the position of a silicon nanocrystal within the structure affects its contribution to the overall PL intensity. We found that in the surface plasmon-polariton regime, the major contribution to the PL intensity comes from the sub-surface silicon nanocrystals. In the waveguide regime, when air slit width is large, the PL is contributed by silicon nanocrystals in depth of the emitting layer. In both cases, the distribution of silicon nanocrystals contribution to the overall PL intensity is highly non-uniform.


This work was supported by the Russian Foundation for Basic Research (No. 15-32-21153 and 16-29-03282). A. Marinins acknowledges support from EU project ICONE (gr. 608099).

%%%%%%%%%%%%%%%%%%%%%%%%%%%%%%%%%%%%%%%%%%%%%%
%

%\bibliography{JAB_library}

\begin{thebibliography}{26}%
\makeatletter
\providecommand \@ifxundefined [1]{%
 \@ifx{#1\undefined}
}%
\providecommand \@ifnum [1]{%
 \ifnum #1\expandafter \@firstoftwo
 \else \expandafter \@secondoftwo
 \fi
}%
\providecommand \@ifx [1]{%
 \ifx #1\expandafter \@firstoftwo
 \else \expandafter \@secondoftwo
 \fi
}%
\providecommand \natexlab [1]{#1}%
\providecommand \enquote  [1]{``#1''}%
\providecommand \bibnamefont  [1]{#1}%
\providecommand \bibfnamefont [1]{#1}%
\providecommand \citenamefont [1]{#1}%
\providecommand \href@noop [0]{\@secondoftwo}%
\providecommand \href [0]{\begingroup \@sanitize@url \@href}%
\providecommand \@href[1]{\@@startlink{#1}\@@href}%
\providecommand \@@href[1]{\endgroup#1\@@endlink}%
\providecommand \@sanitize@url [0]{\catcode `\\12\catcode `\$12\catcode
  `\&12\catcode `\#12\catcode `\^12\catcode `\_12\catcode `\%12\relax}%
\providecommand \@@startlink[1]{}%
\providecommand \@@endlink[0]{}%
\providecommand \url  [0]{\begingroup\@sanitize@url \@url }%
\providecommand \@url [1]{\endgroup\@href {#1}{\urlprefix }}%
\providecommand \urlprefix  [0]{URL }%
\providecommand \Eprint [0]{\href }%
\providecommand \doibase [0]{http://dx.doi.org/}%
\providecommand \selectlanguage [0]{\@gobble}%
\providecommand \bibinfo  [0]{\@secondoftwo}%
\providecommand \bibfield  [0]{\@secondoftwo}%
\providecommand \translation [1]{[#1]}%
\providecommand \BibitemOpen [0]{}%
\providecommand \bibitemStop [0]{}%
\providecommand \bibitemNoStop [0]{.\EOS\space}%
\providecommand \EOS [0]{\spacefactor3000\relax}%
\providecommand \BibitemShut  [1]{\csname bibitem#1\endcsname}%
\let\auto@bib@innerbib\@empty
%</preamble>
\bibitem [{\citenamefont {Noda}\ \emph {et~al.}(2007)\citenamefont {Noda},
  \citenamefont {Fujita},\ and\ \citenamefont {Asano}}]{noda2007spontaneous}%
  \BibitemOpen
  \bibfield  {author} {\bibinfo {author} {\bibfnamefont {S.}~\bibnamefont
  {Noda}}, \bibinfo {author} {\bibfnamefont {M.}~\bibnamefont {Fujita}}, \ and\
  \bibinfo {author} {\bibfnamefont {T.}~\bibnamefont {Asano}},\ }\href@noop {}
  {\bibfield  {journal} {\bibinfo  {journal} {Nature photonics}\ }\textbf
  {\bibinfo {volume} {1}},\ \bibinfo {pages} {449} (\bibinfo {year}
  {2007})}\BibitemShut {NoStop}%
\bibitem [{\citenamefont {Holland}\ and\ \citenamefont
  {Hall}(1985)}]{holland1985waveguide}%
  \BibitemOpen
  \bibfield  {author} {\bibinfo {author} {\bibfnamefont {W.}~\bibnamefont
  {Holland}}\ and\ \bibinfo {author} {\bibfnamefont {D.~G.}\ \bibnamefont
  {Hall}},\ }\href@noop {} {\bibfield  {journal} {\bibinfo  {journal} {Optics
  letters}\ }\textbf {\bibinfo {volume} {10}},\ \bibinfo {pages} {414}
  (\bibinfo {year} {1985})}\BibitemShut {NoStop}%
\bibitem [{\citenamefont {Guzatov}\ \emph {et~al.}(2012)\citenamefont
  {Guzatov}, \citenamefont {Vaschenko}, \citenamefont {Stankevich},
  \citenamefont {Lunevich}, \citenamefont {Glukhov},\ and\ \citenamefont
  {Gaponenko}}]{guzatov2012plasmonic}%
  \BibitemOpen
  \bibfield  {author} {\bibinfo {author} {\bibfnamefont {D.~V.}\ \bibnamefont
  {Guzatov}}, \bibinfo {author} {\bibfnamefont {S.~V.}\ \bibnamefont
  {Vaschenko}}, \bibinfo {author} {\bibfnamefont {V.~V.}\ \bibnamefont
  {Stankevich}}, \bibinfo {author} {\bibfnamefont {A.~Y.}\ \bibnamefont
  {Lunevich}}, \bibinfo {author} {\bibfnamefont {Y.~F.}\ \bibnamefont
  {Glukhov}}, \ and\ \bibinfo {author} {\bibfnamefont {S.~V.}\ \bibnamefont
  {Gaponenko}},\ }\href@noop {} {\bibfield  {journal} {\bibinfo  {journal} {The
  Journal of Physical Chemistry C}\ }\textbf {\bibinfo {volume} {116}},\
  \bibinfo {pages} {10723} (\bibinfo {year} {2012})}\BibitemShut {NoStop}%
\bibitem [{\citenamefont {Tam}\ \emph {et~al.}(2007)\citenamefont {Tam},
  \citenamefont {Goodrich}, \citenamefont {Johnson},\ and\ \citenamefont
  {Halas}}]{tam2007plasmonic}%
  \BibitemOpen
  \bibfield  {author} {\bibinfo {author} {\bibfnamefont {F.}~\bibnamefont
  {Tam}}, \bibinfo {author} {\bibfnamefont {G.~P.}\ \bibnamefont {Goodrich}},
  \bibinfo {author} {\bibfnamefont {B.~R.}\ \bibnamefont {Johnson}}, \ and\
  \bibinfo {author} {\bibfnamefont {N.~J.}\ \bibnamefont {Halas}},\ }\href@noop
  {} {\bibfield  {journal} {\bibinfo  {journal} {Nano letters}\ }\textbf
  {\bibinfo {volume} {7}},\ \bibinfo {pages} {496} (\bibinfo {year}
  {2007})}\BibitemShut {NoStop}%
\bibitem [{\citenamefont {Muskens}\ \emph {et~al.}(2007)\citenamefont
  {Muskens}, \citenamefont {Giannini}, \citenamefont {S{\'a}nchez-Gil},\ and\
  \citenamefont {Gomez~Rivas}}]{muskens2007strong}%
  \BibitemOpen
  \bibfield  {author} {\bibinfo {author} {\bibfnamefont {O.}~\bibnamefont
  {Muskens}}, \bibinfo {author} {\bibfnamefont {V.}~\bibnamefont {Giannini}},
  \bibinfo {author} {\bibfnamefont {J.~A.}\ \bibnamefont {S{\'a}nchez-Gil}}, \
  and\ \bibinfo {author} {\bibfnamefont {J.}~\bibnamefont {Gomez~Rivas}},\
  }\href@noop {} {\bibfield  {journal} {\bibinfo  {journal} {Nano letters}\
  }\textbf {\bibinfo {volume} {7}},\ \bibinfo {pages} {2871} (\bibinfo {year}
  {2007})}\BibitemShut {NoStop}%
\bibitem [{\citenamefont {Vasa}\ \emph {et~al.}(2008)\citenamefont {Vasa},
  \citenamefont {Pomraenke}, \citenamefont {Schwieger}, \citenamefont {Mazur},
  \citenamefont {Kunets}, \citenamefont {Srinivasan}, \citenamefont {Johnson},
  \citenamefont {Kihm}, \citenamefont {Kim}, \citenamefont {Runge},
  \citenamefont {Salamo},\ and\ \citenamefont
  {Lienau}}]{PhysRevLett.101.116801}%
  \BibitemOpen
  \bibfield  {author} {\bibinfo {author} {\bibfnamefont {P.}~\bibnamefont
  {Vasa}}, \bibinfo {author} {\bibfnamefont {R.}~\bibnamefont {Pomraenke}},
  \bibinfo {author} {\bibfnamefont {S.}~\bibnamefont {Schwieger}}, \bibinfo
  {author} {\bibfnamefont {Y.~I.}\ \bibnamefont {Mazur}}, \bibinfo {author}
  {\bibfnamefont {V.}~\bibnamefont {Kunets}}, \bibinfo {author} {\bibfnamefont
  {P.}~\bibnamefont {Srinivasan}}, \bibinfo {author} {\bibfnamefont
  {E.}~\bibnamefont {Johnson}}, \bibinfo {author} {\bibfnamefont {J.~E.}\
  \bibnamefont {Kihm}}, \bibinfo {author} {\bibfnamefont {D.~S.}\ \bibnamefont
  {Kim}}, \bibinfo {author} {\bibfnamefont {E.}~\bibnamefont {Runge}}, \bibinfo
  {author} {\bibfnamefont {G.}~\bibnamefont {Salamo}}, \ and\ \bibinfo {author}
  {\bibfnamefont {C.}~\bibnamefont {Lienau}},\ }\href {\doibase
  10.1103/PhysRevLett.101.116801} {\bibfield  {journal} {\bibinfo  {journal}
  {Phys. Rev. Lett.}\ }\textbf {\bibinfo {volume} {101}},\ \bibinfo {pages}
  {116801} (\bibinfo {year} {2008})}\BibitemShut {NoStop}%
\bibitem [{\citenamefont {Takeda}\ \emph {et~al.}(2006)\citenamefont {Takeda},
  \citenamefont {Nakamura}, \citenamefont {Fujii}, \citenamefont {Miura},\ and\
  \citenamefont {Hayashi}}]{Takeda2006c}%
  \BibitemOpen
  \bibfield  {author} {\bibinfo {author} {\bibfnamefont {E.}~\bibnamefont
  {Takeda}}, \bibinfo {author} {\bibfnamefont {T.}~\bibnamefont {Nakamura}},
  \bibinfo {author} {\bibfnamefont {M.}~\bibnamefont {Fujii}}, \bibinfo
  {author} {\bibfnamefont {S.}~\bibnamefont {Miura}}, \ and\ \bibinfo {author}
  {\bibfnamefont {S.}~\bibnamefont {Hayashi}},\ }\href@noop {} {\bibfield
  {journal} {\bibinfo  {journal} {Appl. Phys. Lett.}\ }\textbf {\bibinfo
  {volume} {89}},\ \bibinfo {pages} {101907} (\bibinfo {year}
  {2006})}\BibitemShut {NoStop}%
\bibitem [{\citenamefont {Dyakov}\ \emph {et~al.}(2016)\citenamefont {Dyakov},
  \citenamefont {Zhigunov}, \citenamefont {Marinins}, \citenamefont
  {Shcherbakov}, \citenamefont {Fedyanin}, \citenamefont {Vorontsov},
  \citenamefont {Kashkarov}, \citenamefont {Popov}, \citenamefont {Qiu},
  \citenamefont {Zacharias}, \citenamefont {Tikhodeev},\ and\ \citenamefont
  {Gippius}}]{PhysRevB.93.205413}%
  \BibitemOpen
  \bibfield  {author} {\bibinfo {author} {\bibfnamefont {S.~A.}\ \bibnamefont
  {Dyakov}}, \bibinfo {author} {\bibfnamefont {D.~M.}\ \bibnamefont
  {Zhigunov}}, \bibinfo {author} {\bibfnamefont {A.}~\bibnamefont {Marinins}},
  \bibinfo {author} {\bibfnamefont {M.~R.}\ \bibnamefont {Shcherbakov}},
  \bibinfo {author} {\bibfnamefont {A.~A.}\ \bibnamefont {Fedyanin}}, \bibinfo
  {author} {\bibfnamefont {A.~S.}\ \bibnamefont {Vorontsov}}, \bibinfo {author}
  {\bibfnamefont {P.~K.}\ \bibnamefont {Kashkarov}}, \bibinfo {author}
  {\bibfnamefont {S.}~\bibnamefont {Popov}}, \bibinfo {author} {\bibfnamefont
  {M.}~\bibnamefont {Qiu}}, \bibinfo {author} {\bibfnamefont {M.}~\bibnamefont
  {Zacharias}}, \bibinfo {author} {\bibfnamefont {S.~G.}\ \bibnamefont
  {Tikhodeev}}, \ and\ \bibinfo {author} {\bibfnamefont {N.~A.}\ \bibnamefont
  {Gippius}},\ }\href {\doibase 10.1103/PhysRevB.93.205413} {\bibfield
  {journal} {\bibinfo  {journal} {Phys. Rev. B}\ }\textbf {\bibinfo {volume}
  {93}},\ \bibinfo {pages} {205413} (\bibinfo {year} {2016})}\BibitemShut
  {NoStop}%
\bibitem [{\citenamefont {Goffard}\ \emph {et~al.}(2013)\citenamefont
  {Goffard}, \citenamefont {G{\'e}rard}, \citenamefont {Miska}, \citenamefont
  {Baudrion}, \citenamefont {Deturche},\ and\ \citenamefont
  {Plain}}]{goffard2013plasmonic}%
  \BibitemOpen
  \bibfield  {author} {\bibinfo {author} {\bibfnamefont {J.}~\bibnamefont
  {Goffard}}, \bibinfo {author} {\bibfnamefont {D.}~\bibnamefont {G{\'e}rard}},
  \bibinfo {author} {\bibfnamefont {P.}~\bibnamefont {Miska}}, \bibinfo
  {author} {\bibfnamefont {A.-L.}\ \bibnamefont {Baudrion}}, \bibinfo {author}
  {\bibfnamefont {R.}~\bibnamefont {Deturche}}, \ and\ \bibinfo {author}
  {\bibfnamefont {J.}~\bibnamefont {Plain}},\ }\href@noop {} {\bibfield
  {journal} {\bibinfo  {journal} {Scientific reports}\ }\textbf {\bibinfo
  {volume} {3}},\ \bibinfo {pages} {2672} (\bibinfo {year} {2013})}\BibitemShut
  {NoStop}%
\bibitem [{\citenamefont {Biteen}\ \emph {et~al.}(2007)\citenamefont {Biteen},
  \citenamefont {Sweatlock}, \citenamefont {Mertens}, \citenamefont {Lewis},
  \citenamefont {Polman},\ and\ \citenamefont {Atwater}}]{Biteen2007}%
  \BibitemOpen
  \bibfield  {author} {\bibinfo {author} {\bibfnamefont {J.}~\bibnamefont
  {Biteen}}, \bibinfo {author} {\bibfnamefont {L.}~\bibnamefont {Sweatlock}},
  \bibinfo {author} {\bibfnamefont {H.}~\bibnamefont {Mertens}}, \bibinfo
  {author} {\bibfnamefont {N.}~\bibnamefont {Lewis}}, \bibinfo {author}
  {\bibfnamefont {A.}~\bibnamefont {Polman}}, \ and\ \bibinfo {author}
  {\bibfnamefont {H.}~\bibnamefont {Atwater}},\ }\href {\doibase
  10.1021/jp074160+} {\bibfield  {journal} {\bibinfo  {journal} {J. Phys. Chem.
  C}\ }\textbf {\bibinfo {volume} {111}},\ \bibinfo {pages} {13372} (\bibinfo
  {year} {2007})}\BibitemShut {NoStop}%
\bibitem [{\citenamefont {Sugimoto}\ \emph {et~al.}(2015)\citenamefont
  {Sugimoto}, \citenamefont {Chen}, \citenamefont {Wang}, \citenamefont
  {Fujii}, \citenamefont {Reinhard},\ and\ \citenamefont
  {Dal~Negro}}]{sugimoto2015plasmon}%
  \BibitemOpen
  \bibfield  {author} {\bibinfo {author} {\bibfnamefont {H.}~\bibnamefont
  {Sugimoto}}, \bibinfo {author} {\bibfnamefont {T.}~\bibnamefont {Chen}},
  \bibinfo {author} {\bibfnamefont {R.}~\bibnamefont {Wang}}, \bibinfo {author}
  {\bibfnamefont {M.}~\bibnamefont {Fujii}}, \bibinfo {author} {\bibfnamefont
  {B.~M.}\ \bibnamefont {Reinhard}}, \ and\ \bibinfo {author} {\bibfnamefont
  {L.}~\bibnamefont {Dal~Negro}},\ }\href@noop {} {\bibfield  {journal}
  {\bibinfo  {journal} {Acs Photonics}\ }\textbf {\bibinfo {volume} {2}},\
  \bibinfo {pages} {1298} (\bibinfo {year} {2015})}\BibitemShut {NoStop}%
\bibitem [{\citenamefont {Gardelis}\ \emph {et~al.}(2016)\citenamefont
  {Gardelis}, \citenamefont {Gianneta},\ and\ \citenamefont
  {Nassiopoulou}}]{gardelis2016twenty}%
  \BibitemOpen
  \bibfield  {author} {\bibinfo {author} {\bibfnamefont {S.}~\bibnamefont
  {Gardelis}}, \bibinfo {author} {\bibfnamefont {V.}~\bibnamefont {Gianneta}},
  \ and\ \bibinfo {author} {\bibfnamefont {A.}~\bibnamefont {Nassiopoulou}},\
  }\href@noop {} {\bibfield  {journal} {\bibinfo  {journal} {Journal of
  Luminescence}\ }\textbf {\bibinfo {volume} {170}},\ \bibinfo {pages} {282}
  (\bibinfo {year} {2016})}\BibitemShut {NoStop}%
\bibitem [{\citenamefont {Nychyporuk}\ \emph {et~al.}(2011)\citenamefont
  {Nychyporuk}, \citenamefont {Zakharko}, \citenamefont {Serdiuk},
  \citenamefont {Marty}, \citenamefont {Lemiti},\ and\ \citenamefont
  {Lysenko}}]{nychyporuk2011strong}%
  \BibitemOpen
  \bibfield  {author} {\bibinfo {author} {\bibfnamefont {T.}~\bibnamefont
  {Nychyporuk}}, \bibinfo {author} {\bibfnamefont {Y.}~\bibnamefont
  {Zakharko}}, \bibinfo {author} {\bibfnamefont {T.}~\bibnamefont {Serdiuk}},
  \bibinfo {author} {\bibfnamefont {O.}~\bibnamefont {Marty}}, \bibinfo
  {author} {\bibfnamefont {M.}~\bibnamefont {Lemiti}}, \ and\ \bibinfo {author}
  {\bibfnamefont {V.}~\bibnamefont {Lysenko}},\ }\href@noop {} {\bibfield
  {journal} {\bibinfo  {journal} {Nanoscale}\ }\textbf {\bibinfo {volume}
  {3}},\ \bibinfo {pages} {2472} (\bibinfo {year} {2011})}\BibitemShut
  {NoStop}%
\bibitem [{\citenamefont {Inoue}\ \emph {et~al.}(2015)\citenamefont {Inoue},
  \citenamefont {Fujii}, \citenamefont {Sugimoto},\ and\ \citenamefont
  {Imakita}}]{inoue2015surface}%
  \BibitemOpen
  \bibfield  {author} {\bibinfo {author} {\bibfnamefont {A.}~\bibnamefont
  {Inoue}}, \bibinfo {author} {\bibfnamefont {M.}~\bibnamefont {Fujii}},
  \bibinfo {author} {\bibfnamefont {H.}~\bibnamefont {Sugimoto}}, \ and\
  \bibinfo {author} {\bibfnamefont {K.}~\bibnamefont {Imakita}},\ }\href@noop
  {} {\bibfield  {journal} {\bibinfo  {journal} {The Journal of Physical
  Chemistry C}\ }\textbf {\bibinfo {volume} {119}},\ \bibinfo {pages} {25108}
  (\bibinfo {year} {2015})}\BibitemShut {NoStop}%
\bibitem [{\citenamefont {Harun}\ \emph {et~al.}(2013)\citenamefont {Harun},
  \citenamefont {Benning}, \citenamefont {Horrocks},\ and\ \citenamefont
  {Fulton}}]{harun2013gold}%
  \BibitemOpen
  \bibfield  {author} {\bibinfo {author} {\bibfnamefont {N.~A.}\ \bibnamefont
  {Harun}}, \bibinfo {author} {\bibfnamefont {M.~J.}\ \bibnamefont {Benning}},
  \bibinfo {author} {\bibfnamefont {B.~R.}\ \bibnamefont {Horrocks}}, \ and\
  \bibinfo {author} {\bibfnamefont {D.~A.}\ \bibnamefont {Fulton}},\
  }\href@noop {} {\bibfield  {journal} {\bibinfo  {journal} {Nanoscale}\
  }\textbf {\bibinfo {volume} {5}},\ \bibinfo {pages} {3817} (\bibinfo {year}
  {2013})}\BibitemShut {NoStop}%
\bibitem [{\citenamefont {Zhigunov}\ \emph {et~al.}(2009)\citenamefont
  {Zhigunov}, \citenamefont {Seminogov}, \citenamefont {Timoshenko},
  \citenamefont {Sokolov}, \citenamefont {Glebov}, \citenamefont {Malyutin},
  \citenamefont {Maslova}, \citenamefont {Shalygina}, \citenamefont {Dyakov},
  \citenamefont {Akhmanov}, \citenamefont {Panchenko},\ and\ \citenamefont
  {Kashkarov}}]{zhigunov2009effect}%
  \BibitemOpen
  \bibfield  {author} {\bibinfo {author} {\bibfnamefont {D.}~\bibnamefont
  {Zhigunov}}, \bibinfo {author} {\bibfnamefont {V.}~\bibnamefont {Seminogov}},
  \bibinfo {author} {\bibfnamefont {V.}~\bibnamefont {Timoshenko}}, \bibinfo
  {author} {\bibfnamefont {V.}~\bibnamefont {Sokolov}}, \bibinfo {author}
  {\bibfnamefont {V.}~\bibnamefont {Glebov}}, \bibinfo {author} {\bibfnamefont
  {A.}~\bibnamefont {Malyutin}}, \bibinfo {author} {\bibfnamefont
  {N.}~\bibnamefont {Maslova}}, \bibinfo {author} {\bibfnamefont
  {O.}~\bibnamefont {Shalygina}}, \bibinfo {author} {\bibfnamefont
  {S.}~\bibnamefont {Dyakov}}, \bibinfo {author} {\bibfnamefont
  {A.}~\bibnamefont {Akhmanov}}, \bibinfo {author} {\bibfnamefont
  {V.}~\bibnamefont {Panchenko}}, \ and\ \bibinfo {author} {\bibfnamefont
  {P.}~\bibnamefont {Kashkarov}},\ }\href {\doibase
  https://doi.org/10.1016/j.physe.2008.08.026} {\bibfield  {journal} {\bibinfo
  {journal} {Physica E: Low-dimensional Systems and Nanostructures}\ }\textbf
  {\bibinfo {volume} {41}},\ \bibinfo {pages} {1006 } (\bibinfo {year}
  {2009})}\BibitemShut {NoStop}%
\bibitem [{\citenamefont {Porto}\ \emph {et~al.}(1999)\citenamefont {Porto},
  \citenamefont {Garcia-Vidal},\ and\ \citenamefont
  {Pendry}}]{porto1999transmission}%
  \BibitemOpen
  \bibfield  {author} {\bibinfo {author} {\bibfnamefont {J.}~\bibnamefont
  {Porto}}, \bibinfo {author} {\bibfnamefont {F.}~\bibnamefont {Garcia-Vidal}},
  \ and\ \bibinfo {author} {\bibfnamefont {J.}~\bibnamefont {Pendry}},\
  }\href@noop {} {\bibfield  {journal} {\bibinfo  {journal} {Physical Review
  Letters}\ }\textbf {\bibinfo {volume} {83}},\ \bibinfo {pages} {2845}
  (\bibinfo {year} {1999})}\BibitemShut {NoStop}%
\bibitem [{\citenamefont {Christ}\ \emph {et~al.}(2003)\citenamefont {Christ},
  \citenamefont {Tikhodeev}, \citenamefont {Gippius}, \citenamefont {Kuhl},\
  and\ \citenamefont {Giessen}}]{Christ2003b}%
  \BibitemOpen
  \bibfield  {author} {\bibinfo {author} {\bibfnamefont {A.}~\bibnamefont
  {Christ}}, \bibinfo {author} {\bibfnamefont {S.~G.}\ \bibnamefont
  {Tikhodeev}}, \bibinfo {author} {\bibfnamefont {N.~A.}\ \bibnamefont
  {Gippius}}, \bibinfo {author} {\bibfnamefont {J.}~\bibnamefont {Kuhl}}, \
  and\ \bibinfo {author} {\bibfnamefont {H.}~\bibnamefont {Giessen}},\
  }\href@noop {} {\bibfield  {journal} {\bibinfo  {journal} {Phys. Rev. Lett.}\
  }\textbf {\bibinfo {volume} {91}},\ \bibinfo {pages} {183901} (\bibinfo
  {year} {2003})}\BibitemShut {NoStop}%
\bibitem [{\citenamefont {Christ}\ \emph {et~al.}(2004)\citenamefont {Christ},
  \citenamefont {Zentgraf}, \citenamefont {Kuhl}, \citenamefont {Tikhodeev},
  \citenamefont {Gippius},\ and\ \citenamefont {Giessen}}]{Christ2004}%
  \BibitemOpen
  \bibfield  {author} {\bibinfo {author} {\bibfnamefont {A.}~\bibnamefont
  {Christ}}, \bibinfo {author} {\bibfnamefont {T.}~\bibnamefont {Zentgraf}},
  \bibinfo {author} {\bibfnamefont {J.}~\bibnamefont {Kuhl}}, \bibinfo {author}
  {\bibfnamefont {S.~G.}\ \bibnamefont {Tikhodeev}}, \bibinfo {author}
  {\bibfnamefont {N.~A.}\ \bibnamefont {Gippius}}, \ and\ \bibinfo {author}
  {\bibfnamefont {H.}~\bibnamefont {Giessen}},\ }\href@noop {} {\bibfield
  {journal} {\bibinfo  {journal} {Phys. Rev. B}\ }\textbf {\bibinfo {volume}
  {70}},\ \bibinfo {pages} {125113} (\bibinfo {year} {2004})}\BibitemShut
  {NoStop}%
\bibitem [{\citenamefont {Zhigunov}\ \emph {et~al.}(2012)\citenamefont
  {Zhigunov}, \citenamefont {Shvydun}, \citenamefont {Emelyanov}, \citenamefont
  {Timoshenko}, \citenamefont {Kashkarov},\ and\ \citenamefont
  {Seminogov}}]{zhigunov2012photoluminescence}%
  \BibitemOpen
  \bibfield  {author} {\bibinfo {author} {\bibfnamefont {D.}~\bibnamefont
  {Zhigunov}}, \bibinfo {author} {\bibfnamefont {N.}~\bibnamefont {Shvydun}},
  \bibinfo {author} {\bibfnamefont {A.}~\bibnamefont {Emelyanov}}, \bibinfo
  {author} {\bibfnamefont {V.~Y.}\ \bibnamefont {Timoshenko}}, \bibinfo
  {author} {\bibfnamefont {P.}~\bibnamefont {Kashkarov}}, \ and\ \bibinfo
  {author} {\bibfnamefont {V.}~\bibnamefont {Seminogov}},\ }\href@noop {}
  {\bibfield  {journal} {\bibinfo  {journal} {Semiconductors}\ }\textbf
  {\bibinfo {volume} {46}},\ \bibinfo {pages} {354} (\bibinfo {year}
  {2012})}\BibitemShut {NoStop}%
\bibitem [{\citenamefont {Biteen}\ \emph {et~al.}(2005)\citenamefont {Biteen},
  \citenamefont {Pacifici}, \citenamefont {Lewis},\ and\ \citenamefont
  {Atwater}}]{biteen2005enhanced}%
  \BibitemOpen
  \bibfield  {author} {\bibinfo {author} {\bibfnamefont {J.~S.}\ \bibnamefont
  {Biteen}}, \bibinfo {author} {\bibfnamefont {D.}~\bibnamefont {Pacifici}},
  \bibinfo {author} {\bibfnamefont {N.~S.}\ \bibnamefont {Lewis}}, \ and\
  \bibinfo {author} {\bibfnamefont {H.~A.}\ \bibnamefont {Atwater}},\
  }\href@noop {} {\bibfield  {journal} {\bibinfo  {journal} {Nano letters}\
  }\textbf {\bibinfo {volume} {5}},\ \bibinfo {pages} {1768} (\bibinfo {year}
  {2005})}\BibitemShut {NoStop}%
\bibitem [{\citenamefont {Tikhodeev}\ \emph {et~al.}(2002)\citenamefont
  {Tikhodeev}, \citenamefont {Yablonskii}, \citenamefont {Muljarov},
  \citenamefont {Gippius},\ and\ \citenamefont {Ishihara}}]{Tikhodeev2002b}%
  \BibitemOpen
  \bibfield  {author} {\bibinfo {author} {\bibfnamefont {S.~G.}\ \bibnamefont
  {Tikhodeev}}, \bibinfo {author} {\bibfnamefont {A.~L.}\ \bibnamefont
  {Yablonskii}}, \bibinfo {author} {\bibfnamefont {E.~A.}\ \bibnamefont
  {Muljarov}}, \bibinfo {author} {\bibfnamefont {N.~A.}\ \bibnamefont
  {Gippius}}, \ and\ \bibinfo {author} {\bibfnamefont {T.}~\bibnamefont
  {Ishihara}},\ }\href@noop {} {\bibfield  {journal} {\bibinfo  {journal}
  {Phys. Rev. B}\ }\textbf {\bibinfo {volume} {66}},\ \bibinfo {pages} {045102}
  (\bibinfo {year} {2002})}\BibitemShut {NoStop}%
\bibitem [{\citenamefont {Whittaker}\ and\ \citenamefont
  {Culshaw}(1999)}]{whittaker99}%
  \BibitemOpen
  \bibfield  {author} {\bibinfo {author} {\bibfnamefont {D.~M.}\ \bibnamefont
  {Whittaker}}\ and\ \bibinfo {author} {\bibfnamefont {I.~S.}\ \bibnamefont
  {Culshaw}},\ }\href@noop {} {\bibfield  {journal} {\bibinfo  {journal} {Phys.
  Rev. B}\ }\textbf {\bibinfo {volume} {60}},\ \bibinfo {pages} {2610}
  (\bibinfo {year} {1999})}\BibitemShut {NoStop}%
\bibitem [{\citenamefont {Moharam}\ \emph {et~al.}(1995)\citenamefont
  {Moharam}, \citenamefont {Gaylord}, \citenamefont {Grann},\ and\
  \citenamefont {Pommet}}]{moharam1995formulation}%
  \BibitemOpen
  \bibfield  {author} {\bibinfo {author} {\bibfnamefont {M.}~\bibnamefont
  {Moharam}}, \bibinfo {author} {\bibfnamefont {T.}~\bibnamefont {Gaylord}},
  \bibinfo {author} {\bibfnamefont {E.~B.}\ \bibnamefont {Grann}}, \ and\
  \bibinfo {author} {\bibfnamefont {D.~A.}\ \bibnamefont {Pommet}},\
  }\href@noop {} {\bibfield  {journal} {\bibinfo  {journal} {JOSA a}\ }\textbf
  {\bibinfo {volume} {12}},\ \bibinfo {pages} {1068} (\bibinfo {year}
  {1995})}\BibitemShut {NoStop}%
\bibitem [{\citenamefont {Li}(1996)}]{li1996use}%
  \BibitemOpen
  \bibfield  {author} {\bibinfo {author} {\bibfnamefont {L.}~\bibnamefont
  {Li}},\ }\href@noop {} {\bibfield  {journal} {\bibinfo  {journal} {JOSA A}\
  }\textbf {\bibinfo {volume} {13}},\ \bibinfo {pages} {1870} (\bibinfo {year}
  {1996})}\BibitemShut {NoStop}%
\bibitem [{\citenamefont {Dyakov}\ \emph {et~al.}(2012)\citenamefont {Dyakov},
  \citenamefont {Zhigunov}, \citenamefont {Hartel}, \citenamefont {Zacharias},
  \citenamefont {Perova},\ and\ \citenamefont
  {Timoshenko}}]{dyakov2012enhancement}%
  \BibitemOpen
  \bibfield  {author} {\bibinfo {author} {\bibfnamefont {S.}~\bibnamefont
  {Dyakov}}, \bibinfo {author} {\bibfnamefont {D.}~\bibnamefont {Zhigunov}},
  \bibinfo {author} {\bibfnamefont {A.}~\bibnamefont {Hartel}}, \bibinfo
  {author} {\bibfnamefont {M.}~\bibnamefont {Zacharias}}, \bibinfo {author}
  {\bibfnamefont {T.}~\bibnamefont {Perova}}, \ and\ \bibinfo {author}
  {\bibfnamefont {V.~Y.}\ \bibnamefont {Timoshenko}},\ }\href@noop {}
  {\bibfield  {journal} {\bibinfo  {journal} {Applied Physics Letters}\
  }\textbf {\bibinfo {volume} {100}},\ \bibinfo {pages} {061908} (\bibinfo
  {year} {2012})}\BibitemShut {NoStop}%
\end{thebibliography}
%\bibliographystyle{apsrev4-1}
\end{document}